\documentclass[preprintnumbers,amsmath,amssymbm,prd]{revtex4}
\usepackage{epsfig}
\usepackage{graphicx}

\begin{document}
\title{Hawking radiation and the Stefan-Boltzmann law:
The effective radius of the black-hole quantum atmosphere}
\author{Shahar Hod}
\affiliation{The Ruppin Academic Center, Emeq Hefer 40250, Israel}
\affiliation{ } \affiliation{The Hadassah Institute, Jerusalem
91010, Israel}
\date{\today}

\begin{abstract}
\ \ \ It has recently been suggested [S. B. Giddings, Phys. Lett. B
{\bf 754}, 39 (2016)] that the Hawking black-hole radiation spectrum
originates from an effective quantum ``atmosphere" which extends
well outside the black-hole horizon. In particular, comparing the
Hawking radiation power of a $(3+1)$-dimensional Schwarzschild black
hole of horizon radius $r_{\text{H}}$ with the familiar
Stefan-Boltzmann radiation power of a $(3+1)$-dimensional flat space
perfect blackbody emitter, Giddings concluded that the source of the
Hawking semi-classical black-hole radiation is a quantum region
outside the Schwarzschild black-hole horizon whose effective radius
$r_{\text{A}}$ is characterized by the relation $\Delta r\equiv
r_{\text{A}}-r_{\text{H}}\sim r_{\text{H}}$. It is of considerable
physical interest to test the general validity of Giddings's
intriguing conclusion. To this end, we study the Hawking radiation
of $(D+1)$-dimensional Schwarzschild black holes. We find that the
dimensionless radii $r_{\text{A}}/r_{\text{H}}$ which characterize
the black-hole quantum atmospheres, as determined from the Hawking
black-hole radiation power and the $(D+1)$-dimensional
Stefan-Boltzmann radiation law, are a decreasing function of the
number $D+1$ of spacetime dimensions. In particular, it is shown
that radiating $(D+1)$-dimensional Schwarzschild black holes are
characterized by the relation
$(r_{\text{A}}-r_{\text{H}})/r_{\text{H}}\ll1$ in the large $D\gg1$
regime. Our results therefore suggest that, at least in some
physical cases, the Hawking emission spectrum originates from
quantum excitations very near the black-hole horizon.
\end{abstract}
\bigskip
\maketitle

\section{Introduction}

The Hawking evaporation process of black holes seems, at first
glance, to be characterized by a non-unitary evolution of quantum
fields in curved spacetimes \cite{Haw}. In particular, according to
Hawking's original analysis, matter fields in a {\it pure} quantum
state may collapse to form a black hole which eventually evaporates
into a {\it mixed} thermal state \cite{Haw}. Since a unitary
temporal evolution of quantum states is one of the cornerstones of
quantum mechanics, it is widely believed that the semi-classical
Hawking radiation spectra of evaporating black holes should be
modified in order to restore quantum unitarity \cite{Gid,Notein}.

What is the characteristic lengthscale associated with these yet
unknown quantum modifications? It is commonly believed that the
Hawking emission spectra of evaporating black holes originate from
quantum excitations in the near-horizon $\Delta r=r-r_{\text{H}}\ll
r_{\text{H}}$ region \cite{Haw,Notein}. It is therefore widely
expected \cite{Notein} that the required quantum modifications of
the semi-classical Hawking radiation spectra would also be
characterized by this relatively short \cite{Noterel} lengthscale
$\Delta r\ll r_{\text{H}}$.

However, in a very interesting work, Giddings \cite{Gid} has
recently suggested that the radiation spectrum of an evaporating
black hole originates from an effective quantum ``atmosphere" which
extends well outside the black-hole horizon. In particular, by
comparing the numerically computed \cite{Page,Noteblock} Hawking
radiation power $P_{\text{BH}}$ of an evaporating
$(3+1)$-dimensional Schwarzschild black hole of horizon radius
$r_{\text{H}}$ with the familiar Stefan-Boltzmann radiation power
$P_{\text{BB}}=\sigma A T^4$ \cite{Notewh} of a $(3+1)$-dimensional
flat space perfect blackbody emitter of radius $r_{\text{A}}$,
Giddings concluded that the source of the Hawking radiation is a
quantum region (the effective black-hole atmosphere) located outside
the black-hole horizon and whose effective radius $r_{\text{A}}$ is
characterized by the relation
\begin{equation}\label{Eq1}
\Delta r\equiv r_{\text{A}}-r_{\text{H}}\sim r_{\text{H}}\  .
\end{equation}
As emphasized in \cite{Gid}, the relation (\ref{Eq1}), which
characterizes the $(3+1)$-dimensional Schwarzschild black hole, is
consistent with the existence of an effective emitting atmosphere
which extends well outside the black-hole horizon.

It is of physical interest to test the general validity of
Giddings's intriguing conclusion (\ref{Eq1}). In particular, one
naturally wonders whether the relation (\ref{Eq1}), which
characterizes the effective quantum atmosphere of the
$(3+1)$-dimensional Schwarzschild black hole, is a generic feature
of {\it all} evaporating black holes?

In order to address this important question, in this paper we shall
study the Hawking radiation powers of $(D+1)$-dimensional
Schwarzschild black holes. In particular, following \cite{Gid} we
shall define the effective radii $r_{\text{A}}(D)$ of the black-hole
quantum atmospheres by equating the Hawking radiation powers of the
$(D+1)$-dimensional black holes with the corresponding
Stefan-Boltzmann radiation powers of flat space perfect blackbody
emitters. Below we shall explicitly show that the dimensionless
radii $r_{\text{A}}/r_{\text{H}},$ which characterize the effective
black-hole quantum atmospheres, are a {\it decreasing} function of
the number $D+1$ of spacetime dimensions. In particular, our results
(to be presented below) suggest that radiating $(D+1)$-dimensional
Schwarzschild black holes are characterized by the relation
$(r_{\text{A}}-r_{\text{H}})/r_{\text{H}}\ll1$ [see Eq. (\ref{Eq17})
below] in the large $D\gg1$ regime.

\section{The Hawking radiation spectra of $(D+1)$-dimensional Schwarzschild black
holes}

We study the Hawking emission of massless scalar fields from
$(D+1)$-dimensional Schwarzschild black holes. The semi-classical
Hawking radiation power for one bosonic degree of freedom is given
by the integral relation \cite{Haw,Page,Noteunit,ZuKa}
\begin{equation}\label{Eq2}
P_{\text{BH}}={{\hbar}\over{2^{D-1}\pi^{D/2}\Gamma(D/2)}}\sum_{j}{\int_0^{\infty}}\Gamma
{{\omega^D}\over{{e^{\hbar\omega/T_{\text{BH}}}-1}}} d\omega\  ,
\end{equation}
where $j$ denotes the angular harmonic indices of the emitted field
modes, and
\begin{equation}\label{Eq3}
T_{\text{BH}}={{(D-2)\hbar}\over{4\pi r_{\text{H}}}}\  ,
\end{equation}
is the semi-classical Bekenstein-Hawking temperature of the black
hole. Here $r_{\text{H}}$ is the horizon radius of the black hole
\cite{Noterh,SchTang}. The dimensionless coefficients
$\Gamma=\Gamma(\omega;j,D)$, which are known as the greybody factors
\cite{Page} of the composed black-hole-field system, quantify the
interaction of the emitted fields with the curved black-hole
spacetime.

\section{The effective radius of the black-hole quantum atmosphere}

As pointed out by Giddings \cite{Gid}, one may define the effective
radius of the black-hole quantum atmosphere by equating the Hawking
radiation power $P_{\text{BH}}$ of the emitting black hole with the
corresponding radiation power $P_{\text{BB}}$ of a flat space
perfect blackbody emitter. The scalar radiation power of a
spherically-symmetric blackbody (BB) of temperature $T$ and radius
$R$ in $D+1$ spacetime dimensions is given by the generalized
Stefan-Boltzmann radiation law \cite{TRCar}
\begin{equation}\label{Eq4}
P_{\text{BB}}=\sigma A_{D-1}(R) T^{D+1}\  ,
\end{equation}
where
\begin{equation}\label{Eq5}
\sigma={{D\Gamma(D/2)\zeta(D+1)}\over{2\pi^{D/2+1}\hbar^D}}
\end{equation}
is the generalized [$(D+1)$-dimensional] Stefan-Boltzmann constant
and
\begin{equation}\label{Eq6}
A_{D-1}(R)={{2\pi^{D/2}}\over{\Gamma(D/2)}}R^{D-1}
\end{equation}
is the surface area of the $(D+1)$-dimensional emitting body.

Following \cite{Gid}, we shall define the effective radius
$r_{\text{A}}$ of the black-hole quantum atmosphere from the
relation \cite{Notecnc,Bekog}
\begin{equation}\label{Eq7}
P_{\text{BH}}(r_{\text{H}},T_{\text{BH}})=P_{\text{BB}}(r_{\text{A}},T_{\text{BH}})\
.
\end{equation}
Taking cognizance of Eqs. (\ref{Eq3}), (\ref{Eq4}), (\ref{Eq5}),
(\ref{Eq6}), and (\ref{Eq7}), one finds
\begin{equation}\label{Eq8}
r_{\text{A}}=\Big[{{\pi}\over{D\zeta(D+1)}}\Big({{4\pi}\over{D-2}}\Big)^{D+1}
\bar P_{\text{BH}}\Big]^{{1}\over{D-1}}\times r_{\text{H}}\
\end{equation}
for the effective radiating radius of the $(D+1)$-dimensional
Schwarzschild black hole, where
\begin{equation}\label{Eq9}
\bar P_{\text{BH}}\equiv P_{\text{BH}}\times
{{r^2_{\text{H}}}\over{\hbar}}\
\end{equation}
is the scaled Hawking radiation power of the black hole.

Our main goal is to determine the functional dependence
$r_{\text{A}}=r_{\text{A}}(D)$ of the effective radius (\ref{Eq8})
of the black-hole quantum atmosphere on the spacetime dimension
$D+1$ of the radiating black hole.

\section{The radius of the black-hole quantum atmosphere: Numerical
and analytical results}

In the present section we shall study the functional dependence
$\bar r_{\text{A}}=\bar r_{\text{A}}(D)$ of the dimensionless ratio
\begin{equation}\label{Eq10}
\bar
r_{\text{A}}\equiv{{r_{\text{A}}-r_{\text{H}}}\over{r_{\text{H}}}}
\end{equation}
which characterizes the effective quantum atmospheres of the
radiating $(D+1)$-dimensional black holes.

\subsection{The $(3+1)$-dimensional case}

The Hawking radiation power of scalar quanta from a
$(3+1)$-dimensional Schwarzschild black hole is given by
\cite{Hiscock}
\begin{equation}\label{Eq11}
P_{\text{BH}}(D=3)=2.976\times 10^{-4}
{{\hbar}\over{r^{2}_{\text{H}}}}\ .
\end{equation}
Substituting (\ref{Eq11}) into (\ref{Eq8}), one finds
\begin{equation}\label{Eq12}
r_{\text{A}}=2.679\times r_{\text{H}}
\end{equation}
for the effective radius of the black-hole quantum atmosphere. This
relation yields
\begin{equation}\label{Eq13}
\bar r_{\text{A}}=1.679
\end{equation}
for the dimensionless radius (\ref{Eq10}) which characterizes the
effective atmosphere of the $(3+1)$-dimensional Schwarzschild black
hole.

\subsection{$(D+1)$-dimensional black holes: Intermediate $D$-values}

In the previous subsection we have seen that the effective quantum
atmosphere of the $(3+1)$-dimensional Schwarzschild black hole is
characterized by the relation $\bar
r_{\text{A}}\equiv(r_{\text{A}}-r_{\text{H}})/r_{\text{H}}=O(1)$
[see Eq. (\ref{Eq13})]. This relation implies that, in the
$(3+1)$-dimensional case, the effective quantum atmosphere of the
black hole extends well outside the black-hole horizon. This finding
supports the interesting conclusion presented in \cite{Gid} for the
$(3+1)$-dimensional case.

We shall now show that the dimensionless radii $\bar
r_{\text{A}}(D)$ [see Eqs. (\ref{Eq8}) and (\ref{Eq10})], which
characterize the effective quantum atmospheres of the radiating
$(D+1)$-dimensional Schwarzschild black holes, are a decreasing
function of the number $D$ of spatial dimensions.

The Hawking radiation powers of scalar quanta from
$(D+1)$-dimensional Schwarzschild black holes were computed
numerically in \cite{ZuKa,CKK}. In Table \ref{Table1} we present,
for intermediate values of the number $D+1$ of spacetime dimensions,
the dimensionless radii $\bar r_{\text{A}}(D)$ [see Eqs. (\ref{Eq8})
and (\ref{Eq10})] which characterize the effective quantum
atmospheres of the $(D+1)$-dimensional radiating black holes. The
results presented in Table \ref{Table1} reveal that the
dimensionless radii $\bar r_{\text{A}}(D)$ of the black-hole quantum
atmospheres are a {\it decreasing} function of the number $D+1$ of
spacetime dimensions.

\begin{table}[htbp]
\centering
\begin{tabular}{|c|c|c|c|c|c|c|c|}
\hline $D+1$ & \ \ 5 \ \ & \ \ 6 \ \ & \ \ 7 \ \ & \ \ 8 \ \ & \ \ 9 \ \ & \ \ 10 \ \ & \ \ 11\ \ \ \\
\hline \ \ $(r_{\text{A}}-r_{\text{H}})/r_{\text{H}}$ \ \ &\ \ \
0.982\ \ \ \ &\ \ \ 0.727\ \ \ \ &\ \ \ 0.590\ \ \ \ &\ \ \ 0.502\ \
\ \
&\ \ \ 0.439\ \ \ \ &\ \ \ 0.391\ \ \ \ &\ \ \ 0.355\ \ \ \ \\
\hline
\end{tabular}
\caption{The dimensionless radii $\bar
r_{\text{A}}(D)\equiv(r_{\text{A}}-r_{\text{H}})/r_{\text{H}}$ which
characterize the effective quantum atmospheres of the
$(D+1)$-dimensional radiating black holes. Here $r_{\text{H}}$ is
the radius of the black-hole horizon and $r_{\text{A}}$ is the
effective radius of the black-hole quantum atmosphere as defined
from (\ref{Eq8}). One finds that the dimensionless radii $\bar
r_{\text{A}}(D)$ of the black-hole quantum atmospheres are a {\it
decreasing} function of the number $D+1$ of spacetime dimensions.}
\label{Table1}
\end{table}

\subsection{$(D+1)$-dimensional black holes: The large-$D$ regime}

In the previous subsection we have used numerical data to
demonstrate the fact that the dimensionless radii $\bar
r_{\text{A}}(D)$, which characterize the quantum atmospheres of the
radiating $(D+1)$-dimensional Schwarzschild black holes, are a
decreasing function of the number $D+1$ of spacetime dimensions. In
the present subsection we shall use analytical tools to show that in
fact $\bar r_{\text{A}}(D)\to0$ in the large $D\gg1$ regime.

The Hawking radiation spectrum of $(D+1)$-dimensional Schwarzschild
black holes is characterized by the frequency distribution
$\omega^{D}/(e^{\hbar\omega/T_{\text{BH}}}-1)$ [see Eq.
(\ref{Eq2})]. One finds that this frequency dependent function has a
peak at
\begin{equation}\label{Eq14}
{{\hbar\omega_{\text{peak}}}\over{T_{\text{BH}}}}=D+W(-De^{-D})\ ,
\end{equation}
where $W(x)$ is the Lambert function and $T_{\text{BH}}$ is the
black-hole temperature as given by (\ref{Eq3}). Substituting
(\ref{Eq3}) into (\ref{Eq14}), one finds the strong inequality
\cite{Hoddd}
\begin{equation}\label{Eq15}
{{\lambda_{\text{peak}}}\over{r_{\text{H}}}}={{8\pi^2}\over{D^2}}[1+O(D^{-1})]\ll1\
\end{equation}
which characterizes the Hawking radiation spectra of the
higher-dimensional Schwarzschild black holes in the large $D\gg1$
regime. The asymptotic large-D relation (\ref{Eq15}) reflects the
fact that the characteristic wavelengths emitted by the
higher-dimensional Schwarzschild black holes in the large $D\gg1$
regime are very {\it short} on the lengthscale $r_{\text{H}}$ set by
the horizon radii of the radiating black holes.

As shown in \cite{Hoddd}, the strong inequality (\ref{Eq15}), which
characterizes the radiating $(D+1)$-dimensional Schwarzschild black
holes in the large $D\gg1$ regime, implies that the corresponding
Hawking emission spectra of these higher-dimensional black holes are
described extremely well by the geometric-optics (short wavelengths)
approximation. In particular, in this large-D regime \cite{Notelag},
the effective radiating radius $r_{\text{A}}$ of the black hole is
determined by the high-energy (short wavelengths) absorptive radius
of the hole \cite{ZuKa,Hoddd,CKK}, which is given by the
$(D+1)$-dimensional geometric-optics relation \cite{ZuKa,Hoddd,CKK}
\begin{equation}\label{Eq16}
{{r_{\text{A}}}\over{r_{\text{H}}}}=\Big({{D}\over{2}}\Big)^{{1}\over{D-2}}\sqrt{{{D}\over{D-2}}}\
.
\end{equation}

From Eq. (\ref{Eq16}) one finds
\begin{equation}\label{Eq17}
{{r_{\text{A}}}\over{r_{\text{H}}}}\to1+O\Big({{\ln
D}\over{D}}\Big)\ \ \ \ \text{for}\ \ \ \ D\gg1\
\end{equation}
in the large-D regime. The relation (\ref{Eq17}) reveals the fact
that, in the large $D\gg1$ regime, the effective radius
$r_{\text{A}}$ of the black-hole quantum atmosphere approaches the
horizon radius $r_{\text{H}}$ of the higher-dimensional black hole.

\section{Summary and Discussion}

In a very interesting paper, Giddings \cite{Gid} has recently
suggested that the Hawking radiation spectrum which characterizes an
evaporating semi-classical black hole originates from an effective
quantum ``atmosphere" which extends well outside the black-hole
horizon. In particular, Giddings has provided evidence that, for a
$(3+1)$-dimensional Schwarzschild black hole, the source of the
Hawking radiation is a quantum region outside the black-hole horizon
whose effective radius $r_{\text{A}}$ is characterized by the
relation $\Delta r\equiv r_{\text{A}}-r_{\text{H}}\sim r_{\text{H}}$
[see Eq. (\ref{Eq1})]. It is certainly of physical interest to test
the general validity of Giddings's intriguing conclusion
(\ref{Eq1}). In particular, one naturally wonders whether the
relation $r_{\text{A}}-r_{\text{H}}\sim r_{\text{H}}$ suggested by
Giddings for the effective quantum atmosphere of the
$(3+1)$-dimensional Schwarzschild black hole is a generic
characteristic of all radiating black holes?

In order to address this important question, we have studied in this
paper the Hawking radiation spectra of $(D+1)$-dimensional
Schwarzschild black holes. Interestingly, it was found that the
dimensionless radii $r_{\text{A}}/r_{\text{H}}$, which characterize
the effective black-hole quantum atmospheres, are a {\it decreasing}
function of the number $D+1$ of spacetime dimensions. In particular,
we have shown that the effective quantum atmospheres of
$(D+1)$-dimensional Schwarzschild black holes are characterized by
the relation [see Eqs. (\ref{Eq10}) and (\ref{Eq17})]
\begin{equation}\label{Eq18}
\bar r_{\text{A}}(D)\to0\ \ \ \ \text{for}\ \ \ \ D\gg1\
\end{equation}
in the large-D regime.

It is interesting to note that the large-D eikonal
(geometric-optics) relation (\ref{Eq16}) provides a remarkably
accurate description of the effective black-hole quantum atmospheres
for {\it all} D-values. In Table \ref{Table2} we present the
dimensionless ratio
$r^{\text{numerical}}_{\text{A}}/r^{\text{analytical}}_{\text{A}}$
between the numerically computed radii
$r^{\text{numerical}}_{\text{A}}(D)$ of the black-hole quantum
atmospheres [see Eq. (\ref{Eq8})] and the analytically predicted
radii $r^{\text{analytical}}_{\text{A}}(D)$ of the large-D eikonal
relation (\ref{Eq16}). One finds a remarkably good agreement between
the accurate (numerically computed) radii (\ref{Eq8}) and the
analytically predicted radii of the eikonal relation (\ref{Eq16})
for all values of the number $D+1$ of spacetime dimensions.

\begin{table}[htbp]
\centering
\begin{tabular}{|c|c|c|c|c|c|c|c|c|}
\hline $D+1$ & \ \ 4 \ \ & \ \ 5 \ \ & \ \ 6 \ \ & \ \ 7 \ \ & \ \ 8 \ \ & \ \ 9 \ \ & \ \ 10 \ \ & \ \ 11\ \ \ \\
\hline \ \
$r^{\text{numerical}}_{\text{A}}/r^{\text{analytical}}_{\text{A}}$ \
\ &\ \ \ 1.031\ \ \ \ &\ \ \ 0.991\ \ \ \ &\ \ \ 0.986\ \ \ \ &\ \ \
0.986\ \ \ \ &\ \ \ 0.988\ \ \ \
&\ \ \ 0.990\ \ \ \ &\ \ \ 0.990\ \ \ \ &\ \ \ 0.992\ \ \ \ \\
\hline
\end{tabular}
\caption{The dimensionless ratio
$r^{\text{numerical}}_{\text{A}}/r^{\text{analytical}}_{\text{A}}$
between the numerically computed radii
$r^{\text{numerical}}_{\text{A}}(D)$ of the black-hole quantum
atmospheres [see Eq. (\ref{Eq8})] and the analytically predicted
radii $r^{\text{analytical}}_{\text{A}}(D)$ of the large-D eikonal
(geometric-optics) relation (\ref{Eq16}). Remarkably, one finds a
good agreement between the accurate (numerically computed) radii
(\ref{Eq8}) of the black-hole quantum atmospheres and the
analytically predicted radii of the eikonal relation (\ref{Eq16})
for all values of the number $D+1$ of spacetime dimensions.}
\label{Table2}
\end{table}

The results presented in this paper reveal that, at least in some
physical cases, the effective radii of the black-hole quantum
atmospheres are characterized by the relation $\bar
r_{\text{A}}\ll1$ [see Eq. (\ref{Eq18})]. This fact suggests that in
these cases, the Hawking radiation originates from quantum
excitations very near the black-hole horizon.

\newpage
\bigskip
\noindent {\bf ACKNOWLEDGMENTS}
\bigskip

This research is supported by the Carmel Science Foundation. I thank
Yael Oren, Arbel M. Ongo, Ayelet B. Lata, and Alona B. Tea for
stimulating discussions.


\begin{thebibliography}{99}

\bibitem{Haw} S. W. Hawking, Commun. Math. Phys. {\bf 43}, 199 (1975).

\bibitem{Gid} S. B. Giddings, Phys. Lett. B {\bf 754}, 39 (2016) [arXiv:1511.08221].

\bibitem{Notein} See \cite{Gid} and references therein.

\bibitem{Noterel} Short on the lengthscale $r_{\text{H}}$ set by the horizon radius of the evaporating black hole.

\bibitem{Noteblock} It should be emphasized that the black-hole Hawking radiation spectrum
is not identical to that of a perfect blackbody emitter. In
particular, the low frequency part of the black-hole emission
spectrum is blocked by the effective curvature potential which
characterizes the exterior black-hole spacetime.

\bibitem{Page} D. N. Page, Phys. Rev. D {\bf 13}, 198 (1976);
D. N. Page, Phys. Rev. D {\bf 14}, 3260 (1976).

\bibitem{Notewh} Here $\sigma$ is the Stefan-Boltzmann proportionality constant and
$A$ is the surface area of the emitting blackbody.

\bibitem{ZuKa} W. H. Zurek, Phys. Rev. Lett. {\bf 49}, 1683 (1982);
D. Page, Phys. Rev. Lett. {\bf 50}, 1013 (1983); P. Kanti, Int. J.
Mod. Phys. A {\bf 19}, 4899 (2004).

\bibitem{Noteunit} We use gravitational units in which $G=c=k_{\text{B}}=1$.

\bibitem{Noterh} The horizon-radius of a $(D+1)$-dimensional Schwarzschild black hole of mass $M$ is
given by \cite{SchTang} $r_{\text{H}}={[{{16\pi M}/{(D-1)\hat
A_{D-1}}}]}^{1/{(D-2)}}$, where $\hat
A_{D-1}={{2\pi^{D/2}}/{\Gamma(D/2)}}$ is the generalized area of a
unit $(D-1)$-sphere.

\bibitem{SchTang} F. R. Tangherlini, Nuova Cimento {\bf 27}, 365 (1963).

\bibitem{TRCar} T. R. Cardoso and A. S. de Castro, Rev. Bras. Ens. Fis. {\bf 27}, 559 (2005).

\bibitem{Notecnc} It is worth emphasizing that our main goal in the present paper is to
use exactly the {\it same} assumptions that were recently made in
\cite{Gid} in order to determine the effective radius
$r_{\text{A}}(D)$ of the black-hole quantum atmosphere and its
functional dependence on the number $D+1$ of spacetime dimensions.
It is worth emphasizing that, in principle, one can fix in Eq.
(\ref{Eq7}) the surface area of the black hole and the blackbody to
be the same and assume that they have different temperatures. In
fact, a similar assumption was used in \cite{Bekog} in order to
determine an effective temperature for the filtered radiation
emitted by a $(3+1)$-dimensional black hole.

\bibitem{Bekog} J. D. Bekenstein, Phys. Rev. Lett. {\bf 70}, 3680
(1993).

\bibitem{Hiscock} B. E. Taylor, C. M. Chambers, W. A. Hiscock, Phys. Rev. D {\bf 58}, 044012
(1998); T. Elster, Phys. Lett. A {\bf 92}, 205 (1983).

\bibitem{CKK} V. Cardoso, M. Cavaglia, and L. Gualtieri, JHEP {\bf 0602}, 021
(2006); R. A. Konoplya and A. Zhidenko, Phys. Rev. D {\bf 82},
084003 (2010); R. Emparan, G. T. Horowitz and R. C. Myers, Phys.
Rev. Lett. {\bf 85}, 499 (2000).

\bibitem{Hoddd} S. Hod, Class. Quant. Grav. {\bf 28}, 105016 (2011)
[arXiv:1107.0797]; S. Hod, Phys. Lett. B {\bf 746}, 22 (2015); S.
Hod, The Euro. Phys. Jour. C {\bf 75}, 329 (2015)
[arXiv:1506.05457].

\bibitem{Notelag} It is worth emphasizing again that the large
$D\gg1$ regime corresponds to the short wavelengths
$\lambda_{\text{peak}}/r_{\text{H}}\ll1$ regime [see Eq.
(\ref{Eq15})].

\end{thebibliography}
\end{document}